\documentstyle[12pt,epsf]{article}
\begin{document}
\baselineskip=19pt
\begin{titlepage}
\begin{flushright}
  KUNS-1659\\[-1mm]
  YITP-00-19\\[-1mm]
  hep-ph/0005009
\end{flushright}

\begin{center}
  \vspace*{1.5cm}
  
  {\large\bf Infrared alignment of SUSY flavor structures}
  \vspace{1cm}
  
  Tatsuo~Kobayashi\footnote{E-mail address:
    kobayash@gauge.scphys.kyoto-u.ac.jp} and
  Koichi~Yoshioka\footnote{E-mail address:
    yoshioka@yukawa.kyoto-u.ac.jp}
  \vspace{5mm}
  
  $^*${\it Department of Physics, Kyoto University
    Kyoto 606-8502, Japan}\\
  $^\dagger${\it Yukawa Institute for Theoretical Physics, Kyoto
    University,\\ Kyoto 606-8502, Japan}
  \vspace{1.5cm}
  
  \begin{abstract}
    The various experimental bounds on flavor-changing interactions
    severely restrict the low-energy flavor structures of soft
    supersymmetry breaking parameters. In this work, we show that with
    a particular assumption of Yukawa couplings, the fermion mass and
    sfermion soft mass matrices are simultaneously diagonalized by
    common mixing matrices and we then obtain an alignment solution
    for the flavor problems. The required condition is generated by
    renormalization group evolutions and achieved at low-energy scale
    independently of high-energy structures of couplings. In this
    case, the diagonal entries of the soft scalar mass matrices are
    determined by gaugino and Higgs soft masses. We also discuss
    possible realizations of this scenario and the characteristic
    sparticle spectrum in the models.
  \end{abstract}
\end{center}
\end{titlepage}

\section{Introduction}

Since the standard model (SM) has been experimentally established
below the weak scale, many kinds of attempts which extend the SM have
been proposed so far. Among these, supersymmetry (SUSY) provides the
most attractive candidate because of its various successful features
of phenomenological applications such as the stability of mass
hierarchy~\cite{hierarchy}, the gauge coupling unification from the
precise electroweak measurements~\cite{unify}, and so on. Moreover it
has been found through intensive study that supersymmetric gauge
theories have theoretically rich phase structures such as dualities
and confinements. These properties have also been applied to try to
construct more fundamental models beyond the SM.

Despite these attractive natures, the supersymmetric extensions of the
SM generally possess a number of questions in addition to not being
observed experimentally~\cite{question}. Some of these are caused by
the existence of supersymmetric partners of the SM fields. They
generally lead new contributions and sometimes disastrous effects,
without some dynamical or symmetrical reasons, to the processes which
should be suppressed. One of these questions is the flavor-changing
neutral current (FCNC) problem. In the SM, the FCNC events are highly
suppressed by the GIM cancellation and its predictions are properly
consistent with the experimental results. In the SUSY standard models,
however, there are new sources of flavor-changing processes from the
squarks and sleptons generation mixings~\cite{FCNC,FCNC2}. These
effects generally tend to overcome the successful SM predictions and
then severely restrict the parameter forms involved.

To satisfy the FCNC requirements, several mechanisms have been
proposed so far: the degeneracy of soft scalar masses~\cite{FCNC}, the
decoupling of first two heavy generations~\cite{decouple}, the
alignment of quark and squark mass matrices~\cite{Align}, and so
on. These mechanisms have been realized from various dynamics of
high-energy (more fundamental) theories and the SUSY breaking effects;
i.e, soft SUSY breaking parameters are thus given at some high-energy
scale. On the other hand, since the FCNC processes are observed in the
low-energy region, the renormalization-group (RG) evolutions between
these two scales should be taken into account. These substantial
effects have been used to obtain the approximate universal forms of
SUSY breaking terms~\cite{RGEuniv} and the heavy sfermion of the first
two generations~\cite{RGEdec}, where specific initial conditions of
parameters at SUSY breaking scale are required to obtain sizable RG
effects for avoiding the FCNC problems.

In this paper we show that with an assumption for Yukawa couplings,
the flavor structure of squarks mass matrix is aligned with the
corresponding Yukawa matrix below the SUSY breaking scale due to the
RG running effects. This infrared alignment occurs provided that the
low-energy values of Yukawa couplings are determined from the infrared
fixed points of RG flow. Only with this simple and model-independent
assumption is it found that the squark squared-mass matrix is
diagonalized by the unitary matrices, which can also diagonalize the
quark mass matrix (i.e., the fixed-point solutions), and then the FCNC
processes are indeed suppressed. The manifestation of alignment is
neither based on any special symmetries nor depends on detailed
structures of high-energy theories. In this case, one may naively
wonder whether small Yukawa couplings compared to gauge couplings can
be realized as the RG fixed points. However, it can be actually
accomplished, for example, by introducing heavy matter fields or by
taking into account the contributions from extra spatial
dimensions. We will discuss these realizations in more detail in a
later section.

To suppress the FCNC processes to an acceptable level, a certain
condition is generally required as in the RG-universality or 
RG-decoupling mechanisms stated above. In the present infrared
alignment scenario, it is the rapid convergence of the fixed-point 
solutions. In the above examples of extra matter and extra dimensions,
we actually obtain strong convergence into the fixed points. Moreover
it has been known that the convergence of soft SUSY breaking
parameters is expressed by the same factor as that of Yukawa
coupling. Thus all properties of the mechanism are described by 
the Yukawa sector only and do not depend on detailed structures of
SUSY breaking parameters at high-energy scale. In the following we
will discuss these phenomena in addition to the characteristic pattern
of sparticle spectrum predicted by the infrared alignment.

\section{RG running of soft terms}
\setcounter{equation}{0}

For later discussions, we first consider the fixed-point structures of
soft breaking parameters in the SUSY standard models, and the
convergence to the fixed points. As stated before, the convergence
behavior is important to discuss the applications of fixed-point
solutions to our low-energy physics. For this purpose we first work in
the model with one gauge and one Yukawa couplings for simplicity. In
realistic models they can be regarded as the $SU(3)_c$ gauge coupling
and the top Yukawa coupling. It is straightforward to extend the
obtained results to the case with more numbers of generation and the
off-diagonal entries in soft SUSY breaking terms. We suppose the
following superpotential:
\begin{eqnarray}
  W = y QuH_u,
\end{eqnarray}
where $Q$, $u$, and $H_u$ denote quark doublet, quark singlet, and
up-type Higgs superfield, respectively. The one-loop beta functions of
the gauge coupling $g$ and Yukawa coupling $y$ are written as
\begin{eqnarray}
  \frac{dg}{dt} &=& \frac{1}{16\pi^2} b g^3, \label{g}\\
  \frac{dy}{dt} &=& \frac{1}{16\pi^2}\, y(a\bar y y-cg^2), \label{y}
\end{eqnarray}
where $t=\ln\mu$ and the coefficients $a$, $b$, and $c$ are $O(1)$
constants expressed in terms of group-theoretical indices. We also 
introduce the soft SUSY breaking terms, a gaugino mass, scalar
soft masses, and a trilinear coupling
\begin{eqnarray}
  -L_{\rm soft} &=& \left(\frac{M}{2}\lambda\lambda +AQuH_u 
    +{\rm h.c.} \right) +Q^\dagger m_Q^2Q +u^\dagger m_u^2 u
  +H_u^\dagger m_{H_u}^2 H_u,
\end{eqnarray}
where $\lambda$ is a gaugino and $Q$, $u$, and $H_u$ are the scalar
components of corresponding chiral superfields. The beta functions for 
these soft SUSY breaking parameters are similarly obtained by the
perturbative calculations of relevant one-loop Feynman
diagrams. However, here we adopt a more convenient and useful way of 
calculations in the light of later discussions in the multi-generation 
case. It is known that the total Lagrangian with the soft SUSY
breaking terms can be written in terms of superfields by introducing
the Grassmannian parameter (external field) $\eta$~\cite{spurion}. With
this formalism, the beta functions of SUSY breaking parameters can
simply be written down by the so-called Grassmannian expansion
method~\cite{GTE}. That is, the supersymmetric couplings $g$ and
$y$ are replaced by
\begin{eqnarray}
  \tilde g^2 &=& g^2 (1+M\eta+\bar M\bar\eta+2M\bar M\eta\bar\eta),\\
  \tilde y &=& y -A\eta+\frac{1}{2}\,y(m_Q^2+m_u^2+m_{H_u}^2)
  \eta\bar\eta,
\end{eqnarray}
and are expanded by the Grassmannian parameter $\eta$. Then one
obtains the beta functions of SUSY breaking parameters from the rigid
SUSY ones (\ref{g},~\ref{y}) as follows:
\begin{eqnarray}
  \frac{dM}{dt} &=& \frac{1}{16\pi^2}\, 2bg^2M,\\
  \frac{dA}{dt} &=& \frac{1}{16\pi^2} \left(3a\bar y yA -cg^2(A-2My)
  \right),\\
  \frac{d\Sigma}{dt} &=& \frac{2}{16\pi^2} \left(a(\bar y y\Sigma
    +\bar AA) -2cg^2M\bar M \right),
\end{eqnarray}
where $\Sigma$ is a sum of the soft scalar masses;
$\Sigma=m_Q^2+m_u^2+m_{H_u}^2$. (In the case with more Yukawa
couplings, the sum $\Sigma$ is defined corresponding to each
Yukawa coupling.) \ From these beta functions, one can find the
infrared stable fixed-point solutions;
\begin{eqnarray}
  \bar yy \;=\; \frac{b+c}{a} g^2,\quad A \;=\; -My,\quad \Sigma \;=\;
  M^2.
  \label{fpsol}
\end{eqnarray}
The fixed-point values are written in terms of the gauge coupling $g$
and the gaugino mass $M$\@. The solutions for SUSY breaking parameters
$A$ and $\Sigma$ can also be derived directly from the solution of
Yukawa coupling $y$ by use of the Grassmannian expansions. Here it is
important to note that the fixed-point solutions for the soft SUSY
breaking parameters do exist {\it only when} the corresponding Yukawa
coupling has a stable fixed point. Therefore in the usual three
generations case that the first two generations have tiny Yukawa
couplings, the soft terms of these generations do not have stable
infrared fixed points like eq.~(\ref{fpsol}).

Before proceeding to the generic three-generation case, we consider
the convergence of couplings to the fixed-point solution
(\ref{fpsol}). For this, let us analytically integrate the RG 
equations. In the present case, i.e, with only one Yukawa coupling,
the results are expressed in a simple form as
\begin{eqnarray}
  X_t \;=\; \xi X_0,\quad (YM)_t \;=\; \xi\,(YM)_0,\quad 
  (ZM^2)_t \;=\; \xi\,(ZM^2)_0,
  \label{int}
\end{eqnarray}
where $\xi\equiv (g(t)/g_0)^{2(1+c/b)}$ and the subscripts 0 imply
they are initial values at some high-energy scale. The combined
couplings $X$, $Y$, and $Z$ are defined by
\begin{eqnarray}
  X &=& \frac{R-1}{R},\\
  Y &=& \frac{S-1}{R}-\left(1+\frac{c}{b}\right)\frac{R-1}{R},\\
  Z &=& \frac{T-1}{R}-\frac{(S-1)(\bar S-1)}{R} \nonumber \\
  &&-\left(1+\frac{c}{b}\right)\frac{S-1}{R}
  -\left(1+\frac{c}{b}\right)\frac{\bar S-1}{R}
  +\left(1+\frac{c}{b}\right)\left(\frac{c}{b}\right)\frac{R-1}{R},
\end{eqnarray}
where
\begin{eqnarray}
  R \;=\;\frac{a}{b+c}\,\frac{\bar yy}{g^2},\qquad S \;=\;
  \frac{-A}{My},\qquad T \;=\; \frac{\Sigma}{M\bar M}.
\end{eqnarray}
The fixed-point solutions (\ref{fpsol}) correspond to $X=Y=Z=0$ (in
other words, $R=S=T=1$). It can be seen from the integrated forms
(\ref{int}) that the parameter $\xi$ denotes the convergence factor,
that is, the rate at which the couplings approach their fixed-point
values. A sufficiently small value of $\xi$ assures that the couplings
converge to the infrared fixed points rapidly. As will be seen below,
it depends on this smallness of $\xi$ how the FCNC processes, which
involve the off-diagonal elements of soft mass matrix, can be
suppressed. Interestingly enough, all the RG fixed points including
the soft SUSY breaking parameters are controlled by the single
convergence factor $\xi$. Thus the FCNC problems of SUSY breaking
parameters may be studied only from the rigid SUSY sector, and the
initial values of these couplings are irrelevant to discussions. The
factor $\xi$ is expressed only by the gauge interaction part. In the
four-dimensional theories, the smallness of $\xi$ requires that the
gauge couplings should be large at high-energy scale~\cite{ANFfp}. In
this type of model, various phenomenological aspects, e.g., the
Yukawa coupling textures, have been investigated~\cite{ANF}.

\section{Infrared alignment of flavor structures}
\setcounter{equation}{0}

Now we further investigate the generation structures of soft
breaking terms, from which one can discuss the flavor problems in
supersymmetric theories. We consider the case where the RG evolutions
are dominated by one gauge coupling $g$ for simplicity. The
superpotential takes the following form:
\begin{eqnarray}
  W &=& y^{ij} Q_iu_j H_u,
\end{eqnarray}
where $i,j$ are generation indices, $i,j=1,2,3$. Here we consider the
up-quark part only, but the results are easily applied to include the
down-quark sector. The soft SUSY breaking parameters, $m_Q^2$,
$m_u^2$, and $A$, are similarly extended to $3\times 3$ complex-valued
matrices. The scalar squared-mass matrices are hermitian by
definition. We now study the renormalization-group effects on the soft
breaking parameters. A convenient way to see the behaviors of SUSY
breaking terms under RG evolutions is the Grassmannian expansion
method as we have used before. With this method, it is not necessarily
needed to examine the RG equations explicitly. In the multi-generation
case, the expansion of the rigid SUSY couplings are given by
\begin{eqnarray}
  \tilde g^2 &=& g^2 (1+M\eta+\bar M\bar\eta+2M\bar M\eta\bar\eta),
  \label{expg} \\[1mm]
  \tilde y^{ij} &=& y^{ij} -A^{ij}\eta+\frac{1}{2}
  \left( (m_Q^2y)^{ij}+(ym_u^{2\,{\rm T}})^{ij}+y^{ij}m_{H_u}^2
  \right) \eta\bar\eta.
  \label{expy}
\end{eqnarray}

Let us suppose that all the Yukawa couplings $y^{ij}$ are determined
by the infrared fixed-point values. This is the key assumption to
obtaining an alignment of quark-squark flavor structures from RG
evolution. The condition implies that the forms of the fixed-point
solutions for SUSY breaking terms can be read from the above
Grassmannian expansions. It should be noted that the expansion method
can be applied only to the quantities concerned with the
renormalization properties. We therefore apply it to the present
situation that the couplings are determined from the anomalous
dimensions of matter fields. Of course, the following argument of
alignment does not hold for generic Yukawa solutions obtained from
other mechanisms.

The fixed-point solution of Yukawa couplings may generally be given 
by $y^{ij}=c^{ij}f(g)$, or more exactly,
\begin{eqnarray}
  (yy^\dagger)^i_j \;=\; C^i_j F(g^2),\qquad
  (y^\dagger y)^i_j \;=\; \bar C^i_j F(g^2).
  \label{fp}
\end{eqnarray}
The function $F$ does not carry generation indices because the gauge
interaction is flavor blind. In the simplest case, $F$ becomes
$F\propto g^2$ but may have a more complicated form for other
solutions such as quasi fixed points.\footnote{The quasi fixed-point
  solutions may be obtained analytically even in multi-Yukawa coupling 
  cases~\cite{analy}.} The constant matrices $C$ and $\bar C$ take
different forms according to the properties of the doublet and singlet
quarks $Q$ and $u$. These matrices are hermitian by definition and are
diagonalized as
\begin{eqnarray}
  C' &=& V_L^\dagger\,C\,V_L \;=\; V_R^\dagger\,\bar C\,V_R \;=\;
  \mbox{diagonal}
\end{eqnarray}
with two unitary matrices $V_L$ and $V_R$. The Yukawa coupling (the
fixed-point solution $c^{ij}$) is also diagonalized by these matrices
as $y'=V_L^\dagger\,y\,V_R=$ diagonal. Here, the prime means a matrix
in the Yukawa-diagonal basis.

Let us perform the Grassmannian expansions about the fixed-point
solution of $yy^\dagger$. From the expansion (\ref{expy}), we obtain
\begin{eqnarray}
  \widetilde{yy^\dagger} &=& yy^\dagger -Ay^\dagger\eta 
  -yA^\dagger \bar\eta +\left[ \frac{1}{2}y\left(y^\dagger m_Q^2
      +m_u^{2*} y^\dagger +y^\dagger m_{H_u}^2 \right)
  \right. \nonumber \\
  &&\hspace*{15ex}\left. +\frac{1}{2}
    \left( m_Q^2y+ym_u^{2\,{\rm T}}+ym_{H_u}^2\right) y^\dagger
    +AA^\dagger \right] \eta\bar\eta.
\end{eqnarray}
Here and hereafter we omit the generation indices for
simplicity. In the basis that the Yukawa matrix is diagonal, the
solution can be rewritten as
\begin{eqnarray}
  \widetilde{y'{y'}^\dagger} &=& y'{y'}^\dagger -A'{y'}^\dagger\eta
  -y'{A'}^\dagger \bar\eta +\left[ \frac{1}{2}(y'{y'}^\dagger m_Q^{2'}
    +m_Q^{2'}y'{y'}^\dagger) \right. \nonumber \\
  &&\hspace*{20ex}\left. +y'm_u^{2'}{y'}^\dagger +y'{y'}^\dagger
    m_{H_u}^2 +A'{A'}^\dagger \right]\eta\bar\eta,
  \label{expyy}
\end{eqnarray}
where the redefined couplings $A'$, $m_Q^{2'}$, and $m_u^{2'}$ are
given by
\begin{eqnarray}
  A' \;=\; V_L^\dagger\, A\, V_R,\qquad 
  m_Q^{2'} \;=\; V_L^\dagger\, m_Q^2\, V_L,\qquad 
  m_u^{2'} \;=\; V_R^\dagger\, m_u^{2\,{\rm T}}\, V_R.
\end{eqnarray}
These are nothing but the soft SUSY breaking parameters in the
so-called super-CKM basis (at the scale where the fixed points are 
realized). Therefore, if the off-diagonal elements of $A'$,
$m_Q^{2'}$, and $m_u^{2'}$ vanish, the supersymmetric FCNC processes
are suppressed. With the expansions (\ref{expg}) and (\ref{expyy}),
the linear term in $\eta$ of eq.~(\ref{fp}) reads
\begin{eqnarray}
  A'{y'}^\dagger &=& -C'Mg^2\frac{dF}{dg^2}.
\end{eqnarray}
It is easily found that since the matrices $y'$ and $C'$ are diagonal,
the off-diagonal entries in $A'$ vanish. On the other hand, the
diagonal elements, i.e., the eigenvalues of the matrix $A$ are given
by use of the leading term of (\ref{fp}),
\begin{eqnarray}
  {A'}^{ii} &=& -{y'}^{ii}M\,g^2\frac{d\ln F}{dg^2},\qquad
  (i=1,2,3).
  \label{A}
\end{eqnarray}
Thus the trilinear coupling $A$ is completely proportional to the
Yukawa couplings on the fixed point. Again note that this
proportionality is obtained from the assumption that the Yukawa
couplings are determined by fixed-point values. This result is
different from the infrared universality~\cite{JJRA} in which the
couplings are required to satisfy specific conditions by some symmetry
and other theoretical assumptions, or there is no relation between the
Yukawa and SUSY breaking couplings. In the present case, the
coefficient matrices $C$ and $\bar C$ do not necessarily satisfy any
restricted conditions. The Yukawa couplings, trilinear terms, and soft
scalar masses as seen below, take arbitrary forms in the Lagrangian
but can be simultaneously diagonalized by the same mixing
matrices. Note also that the $CP$ phases of trilinear couplings are no
longer free and fixed by that of gaugino mass and Yukawa couplings.

For the soft scalar masses, it is also easily found from the
$\eta\bar\eta$ terms of (\ref{fp}) that the off-diagonal elements
should satisfy
\begin{eqnarray}
  (m_Q^{2'})^i_j \;=\; (m_u^{2'})^i_j \;=\; 0, \qquad (i\neq j)
\end{eqnarray}
provided that all the eigenvalues of the Yukawa matrix are not equal
to each other; $|{y'}^{ii}|^2\neq |{y'}^{jj}|^2$ ($i\neq j$), e.g., if
the mass hierarchy is properly realized. Here we have used the
matrices $y'$ and $A'$ which are diagonal. In addition, the diagonal
elements are given by
\begin{eqnarray}
  (m_Q^{2'})^i_i \,+\, (m_u^{2'})^i_i \,+\, m_{H_u}^2 &=& M^2
  \frac{d}{dg^2}\Bigl(g^4\frac{d}{dg^2}\ln F\Bigr). \qquad (i=1,2,3)
  \label{sum}
\end{eqnarray}
It is interesting to note that this expression does not completely
depend on the model-dependent constants $c^{ij}$. In the end, we find
that the supersymmetric FCNC problems are avoided on the fixed 
point. Namely, the quark and squark mass matrices are simultaneously
diagonalized (``infrared alignment'') if the Yukawa couplings are
determined by RG fixed-point solutions. Exactly speaking, the
alignment suppresses the gaugino-mediated processes only. The
higgsino-mediated processes still exist as a supersymmetric
counterpart of the GIM-suppressed processes in the SM\@. These
contributions are, however, small due to the weak couplings in the
vertex.

As we have shown, the soft SUSY breaking parameters are not universal
in the Lagrangian and may have rather generic structures. This is due
to the remaining freedom of the mixing matrices $V_L$ and $V_R$ and
also the fact that the soft mass eigenvalues are restricted only by
several relations (\ref{sum}). In spite of these facts, interestingly
enough, the FCNC problem is actually settled by quark-squark
alignment. The mixing matrices, i.e., the solution $c^{ij}$ is fixed
in a model-dependent way but their detailed values do not affect the
alignment of soft terms and moreover do not disturb the scalar mass
spectrum either. It is clear from the above derivation that the
alignment occurs in the sector where Yukawa couplings really converge
to their fixed points. For example, when only eigenvalues are fixed,
the flavor structures are not aligned between quarks and squarks
although the relations like (\ref{A}) and (\ref{sum}) are obtained and
somewhat restrict scalar mass matrix forms.

\section{Illustrative models}
\setcounter{equation}{0}

Here we will comment on possible dynamics that may actually realize
the infrared alignment mechanism. As argued above, the mechanism
requires that Yukawa couplings, in particular even small ones, are
reproduced as their fixed-point values. This situation naively cannot
be accomplished in the usual supersymmetric standard models. That is,
the RG evolutions produce only $O(1)$ fixed-point values of Yukawa
couplings because the gauge couplings, which drive the Yukawa
couplings are also $O(1)$. This problem is, however, avoided in some
extensions of the SM.

The first example is to add extra (vector-like) matter fields to the
usual three-generation models. The mass scales of these extra matter
fields are experimentally bounded from below and should be very
large. With the heavy masses, the mixing Yukawa couplings between
extra matter and our three-generation fields could be $O(1)$ and
determined by fixed points of RG running above the heavy mass
scales. Furthermore such heavy masses also depend on the fixed-point 
predictions. If these mass terms are generated from Yukawa couplings,
the vacuum expectation values of relevant Higgs fields should be
assumed as in the electroweak symmetry breaking. As an example, in
case of three extra generations, the $6\times 6$ mass matrix may take
the following form:
\begin{eqnarray}
\left(
{\def\arraystretch{1.3}
  \begin{array}{c|c}
    & O(1)\,m \\ \hline
    O(1)\,m & O(1)\,M
  \end{array}}
\right),
\end{eqnarray}
where $m$ and $M$ are light (electroweak) and heavy mass scales, 
respectively, and $O(1)$'s denote the fixed-point values of Yukawa
couplings. In this case, the small Yukawa couplings of the first two
generations are explained by the heavy mass suppressions like the
seesaw mechanism and are realized as the fixed points. Thus the
condition for alignment is satisfied and in the low-energy effective
theory below $M$, the FCNC processes are suppressed.

Another important point to add extra matter fields is that, in
general, it leads large gauge coupling constants at high energy. As a
consequence, the convergence factor $\xi$ defined in eq.~(\ref{int})
becomes very small and the Yukawa and SUSY breaking couplings are
driven into their fixed points very rapidly~\cite{LR,KY}. This strong
convergence is surely required for the alignment to be achieved, i.e.,
to obtain enough suppressions of the off-diagonal elements of SUSY
breaking parameters in the super-CKM basis. It is interesting that
adding extra matter can provide two conditions required for the
FCNC problem in one effort.

The second example we present is the models with extra spatial
dimensions beyond the usual four dimensions. The quark and lepton
chiral superfields as well as the gauge multiplets could be supposed
to propagate through the extra dimensions. When the compactification
scale $M_c$ of extra dimensions is smaller than the cutoff $\Lambda$,
the effects of RG evolutions between these two scales are enhanced by
the contributions from numbers of Kaluza-Klein excited
modes~\cite{power}. For instance, the beta function of Yukawa 
coupling $y$ is given by
\begin{eqnarray}
  \frac{dy}{dt} &=& \frac{1}{16\pi^2}\,y
  \left[a\biggl(\frac{\Lambda}{\mu}\biggr)^{\delta_y}\bar yy
  -c\biggl(\frac{\Lambda}{\mu}\biggr)^{\delta_g}g^2\right],
\end{eqnarray}
where $O(1)$ coefficients $a$ and $c$ contain group-theoretical
indices and the volume factors originated from the phase-space
integral of Kaluza-Klein modes. Here we have neglected the logarithmic 
terms from the contributions of ordinary four-dimensional
particles. The gauge beta functions are also written in a similar
way. Compared to the four-dimensional case (\ref{y}), the beta
functions are amplified by the power 
factors $(\Lambda/\mu)^x\,(\gg 1)$. Roughly speaking, these powers
just correspond to the number of Kaluza-Klein modes propagating in the
loop diagrams. The integer exponents $\delta_g$ and $\delta_y$ are the
largest gauge and Yukawa contributions, respectively, to the 
anomalous dimensions of matter fields. Of course, the four-dimensional
results are recovered by taking the limit $\delta\to 0$. These
exponents are determined once we fix the configuration of relevant
fields in the extra dimensions. Therefore $\delta_g$ and $\delta_y$
can generally take different values for various Yukawa couplings.

With the enhancement of RG evolutions, the suppressions of Yukawa
couplings are established while the gauge couplings are order
one. The hierarchy between generations can be generated by the
difference of exponents $\delta_x$. More interestingly, these
suppressed Yukawa couplings are actually determined by the infrared
fixed points and do not depend on high-energy input parameters. After
all, it is found that the fermion-sfermion alignment occurs at the
compactification scale $M_c$. In higher-dimensional models, two types
of fixed-point scenarios can be realized by taking suitable field
configurations in the extra dimensions: the quasi fixed
point~\cite{quasi} and the Pendleton-Ross type fixed point~\cite{BKNY}
scenarios. Especially in the case of Pendleton-Ross fixed points, even
the smallness of CKM matrix elements may be explained and the
supersymmetric FCNC processes are completely suppressed by the
extra-dimensional mechanism.

Let us comment on the convergence factor $\xi$ in this case,
too. Since the couplings have power-running behaviors and the running
region may be narrower than the four-dimensional case, one should also
require the strong convergence in order to have enough FCNC
suppressions. The rate of suppression $\xi$ is now given by
\begin{eqnarray}
  \xi &=& \exp\biggl[\int^t_0 dt'\,\frac{c}{8\pi^2}
    \biggl(\frac{\Lambda}{\mu'}\biggr)^{\delta_g}g^2\biggr]
  \biggl(\frac{g^2(t)}{g^2(0)}\biggr).
\end{eqnarray}
This expression certainly reproduces the four-dimensional value (see
section 2) in the limit $\delta_g\to 0$. From this equation, we can 
see that there are two distinct ways to obtain a small value 
of $\xi$~\cite{BKNY}. One is realized in asymptotically non-free gauge
theories. This is clearly seen when the effect from the gauge
anomalous dimension is dominant in the RG evolutions. The situation is
almost similar to the ordinary four-dimensional
cases~\cite{ANFfp}. The other possibility is essentially due to the
existence of extra dimensions. That is, when the gauge contributions
in the matter anomalous dimensions (the $\delta_g$ term) govern the RG
equations, the suppression factor $\xi$ becomes very small. This
result holds even in asymptotically free gauge theories such as the
minimal supersymmetric standard model. Then we may not necessarily
require large gauge coupling constants at high energy unlike the
four-dimensional cases and have many possibilities of model building.

\section{Sparticle spectrum}
\setcounter{equation}{0}

The present alignment realization leads to a rather characteristic
scalar mass spectrum in the models. Here we will briefly discuss these
spectra predicted by the eigenvalue relations of SUSY breaking
parameters (\ref{A}) and (\ref{sum}). The relations give the boundary
conditions of parameters in the low-energy effective theories at the
scale at which the alignment occurs. It is noted that the relations
also hold for the first and second generations. This is due to the
fact that we have assumed that the Yukawa couplings of these
generations are determined from the fixed points.

First, as noted before, the soft scalar mass terms are generally not
universal forms but interestingly, the averaged squark masses are
equal to each other and are given by the gluino and Higgs soft mass
parameters. This is true for the first two generations even when the 
RG-running effects are included because of the small Yukawa
couplings. On the other hand, the averaged mass of the third
generation may differ from the others due to the large top (and
possibly bottom) Yukawa effects. Moreover for the third generation,
there could be a large splitting between two mass eigenvalues by a
large contribution from the trilinear coupling. For the first two
generations, the $A$ parameters are small (see the boundary condition
(\ref{A})).

The results of the previous sections are applied to the down-quark
part and we here assume that the alignment also takes place in the
down sector. In this case, the soft masses relations such as
(\ref{sum}) lead to more interesting features when taking into account
the electroweak symmetry breaking. For example, minimizing the Higgs
potential requires the 
condition, $m_{H_d}^2-m_{H_u}^2=-(m_A^2+M_Z^2)\cos2\beta$, where $m_A$
and $M_Z$ are the masses of the neutral pseudo-scalar Higgs and $Z$
boson. Together with the mass relations, the symmetry breaking
condition is written as
\begin{eqnarray}
  \tilde m_u^2-\tilde m_d^2 &=& -(m_A^2+M_Z^2)\cos2\beta,
\end{eqnarray}
where $\tilde m_x^2$ denotes the averaged squark masses. The
right-hand side is positive since the condition $\cos2\beta>0$ is
inconsistent with the observed quark mass patterns. We thus find that
the up-type squarks are necessarily heavier than the down-type
ones. This characteristic spectrum is different from the usual cases,
e.g., with the universal soft masses generated in the supergravity
models. In that case, the up-type squarks, particularly in the third
generation, become lighter than the down squarks at the electroweak
scale because of the RG effects induced by large Yukawa couplings.

In the above we have considered that the alignment occurs at the
electroweak scale. However, the conditions may generally be achieved
at some higher scale. In the realistic models discussed in the
previous section, this scale corresponds to the heavy mass scale of
extra matter fields or the compactification scale of extra
dimensions. For definiteness, we suppose that below these scales
we have the minimal supersymmetric standard model. Then the difference
of the low-energy stop and sbottom averaged 
masses $\Delta \tilde m^2\,(=\tilde m_t^2-\tilde m_b^2)$ is evaluated
once one fixes the value of $\tan\beta$.
\begin{figure}[htbp]
  \begin{center}
    \leavevmode
    \epsfxsize=11cm \ \epsfbox{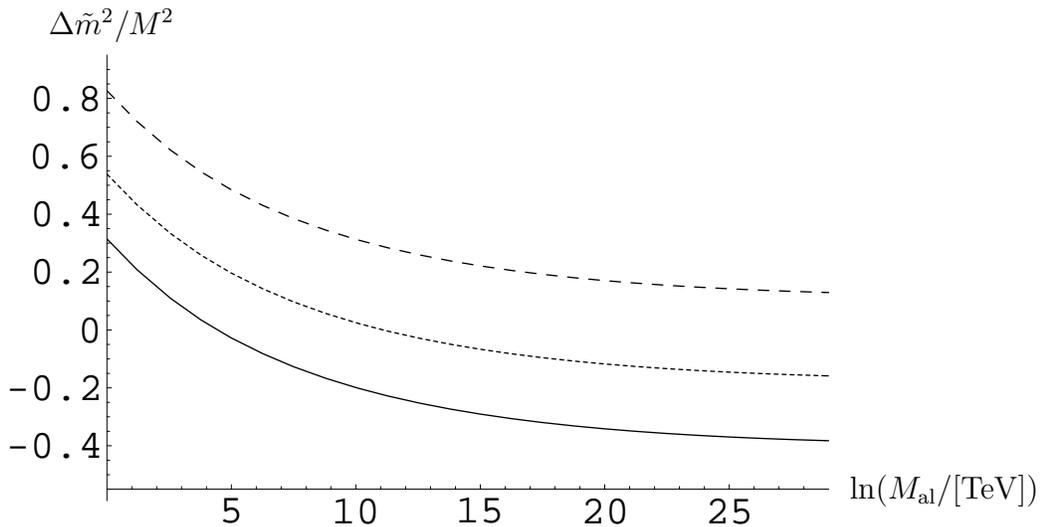}
    \put(8,14){$\ln(M_{\rm al}/[{\rm TeV}])$}
    \put(-295,190){$\Delta\tilde m^2/M^2$}
    \caption{Typical mass difference of the up and down squarks as a
    function of $M_{\rm al}$ at which scale the alignment occurs. The
    solid, dotted and dashed lines correspond to $m_A=300$, 400, and
    500 GeV, respectively. We take $\tan\beta=3$ and the gluino mass
    $M=500$ GeV at 1 TeV.}
  \label{diff}
  \end{center}
\end{figure}
In fig.~\ref{diff}, we show a typical behavior of the mass 
difference $\Delta \tilde m^2$ as a function of $M_{\rm al}$ at which
scale the alignment occurs, i.e., the soft masses relations are
imposed. Here we take $\tan\beta=3$ and the gluino mass $M=500$~GeV 
at 1~TeV, but the mass difference behavior is almost independent 
of $M$. It is found from this figure that the stop mass can be smaller
than that of sbottom as the scale of new physics $M_{\rm al}$
increases. In other words, if the squark masses are measured in the
future, the new physics scale could be determined. As for the first
and second generations, the up-type squarks are always heavier than
the down-type ones even below the scale $M_{\rm al}$. Generally, the
detailed information above the threshold $M_{\rm al}$ is lost, such
as in the present models using infrared fixed points. However, the RG
invariant relations among the soft SUSY breaking parameters~\cite{KY}
may be helpful in speculating on the SUSY breaking mechanism at
high-energy scale.

We also comment on the masses of neutralino as a possible candidate of
the lightest SUSY particle (LSP) in this scenario. As above, if we
take into account the electroweak symmetry breaking properly, the
supersymmetric Higgs mass parameter $\mu$ is written 
as $\mu^2\simeq -m_{H_u}^2-1/2M_Z^2$ for not too small a value 
of $\tan\beta$. Suppose that the alignment is realized at high energy
and then the soft mass relations are imposed at that scale, the
low-energy value of $\mu$-parameter is given by
\begin{eqnarray}
  \mu^2 &\simeq& \tilde m_t^2-xM^2-\frac{1}{2}M_Z^2,
  \label{mu}
\end{eqnarray}
where the coefficient $x$ denotes the RG-running effect ($x(0)=1$). It
takes the value $0.8-1.0$ depending on the alignment scale, but is
almost independent of $\tan\beta$ and the initial value of $M$. The
averaged mass $\tilde m_t^2$ can vary with different initial ratio of
two soft scalar masses because only the sum of them is known at the
boundary. The value of the $\mu$-parameter roughly gives the
Higgsino-like neutralino mass. Comparing eq.~(\ref{mu}) with the bino
mass, we find that may be a candidate for the LSP\@. If one assume the
universal gaugino masses at some high-energy scale, the bino mass
becomes lighter than the gluino and then it is found from
eq.~(\ref{mu}) that the LSP is bino in almost all parameter
space. However more involved boundary condition of gaugino masses
could naturally induce the Higgsino-like neutralino LSP\@. Of course
this naive analysis does not include the lepton sector, gravitino,
etc., and clearly more exact treatment should be performed, for
example, by embedding the models in grand unified frameworks.

Finally, the FCNC problems in the case of such a higher-scale
alignment should be discussed. As for the gluino-mediated FCNC
processes, it is clear that they are completely suppressed even in the
low-energy region because of the alignment property. This is different
from the case of universal soft SUSY breaking terms. In that case, the
Yukawa matrices are not diagonal in general. Then the off-diagonal
elements of the soft mass-squared matrices are induced under the RG
evolutions and would lead the FCNC problems. On the other hand, in the
alignment mechanism, the Yukawa and soft mass matrices are
simultaneously diagonalized and then the radiative corrections do not
cause the FCNC processes at low energy. However the chargino-mediated
diagrams still exist even in the present case. Although these
processes are suppressed by small coupling constants, they would lead
some bounds for degeneracy between the diagonal elements of the soft
mass matrix of left-handed-type squarks. If that is the case, the soft
mass relations for the first two generations could restrict the scalar
spectra in the models. Notice that one may easily avoid these bounds
by arising the scalar masses (and gaugino masses due to the boundary
conditions from the alignment). A detailed analyses of sparticle mass
patterns including the LSP and also other problems, e.g., the charge
and color breaking minima involving generation mixing couplings, will
be discussed elsewhere.

\section{Summary}
\setcounter{equation}{0}

In this paper we have shown that the quark mass and the squark
squared-mass matrices are diagonalized simultaneously in the case
where Yukawa couplings (quark masses) are determined from the
anomalous dimensions, i.e., the fixed-point solutions of RG
equations. This result provides a natural alignment solution of the
supersymmetric FCNC problems. On the fixed point, the soft SUSY
breaking parameters are not universal but can take arbitrary forms in
the Lagrangian. This is because: (i) the soft mass eigenvalues are
determined only from several relations and (ii) the flavor structure
of Yukawa couplings, i.e., the diagonalizing matrices of the Yukawa
matrix, are not fully fixed by observation. In spite of these
unconstrained mass patterns, the FCNC processes are actually
suppressed at low energy.

We have also discussed several possible dynamics realizing the
infrared alignment. As examples, we have presented the models with
extra heavy fields or with extra spatial dimensions beyond the
SM\@. One of the important points for applying the fixed-point
solution to the FCNC problem is that the SUSY breaking parameters
actually converge to their fixed-point values at an accurate level. We
have shown that this strong convergence can be naturally obtained in
these models.

Furthermore the present alignment also restricts the eigenvalues of
soft SUSY breaking couplings even in the first two generations. The
scalar trilinear couplings are proportional to the corresponding
Yukawa couplings and cannot have new $CP$-violating phases. In
addition, the eigenvalues of squark squared-mass matrices for all
generations are determined by gaugino masses from sum rules. These
results give a peculiar pattern of sparticle mass spectrum. For
example, the up-type squarks are generally heavier than the down-type
squarks. Considering the mass relations as boundary conditions at
some high-energy scale, we also discuss the typical RG-running effects
below that scale and possible candidates of the lightest SUSY
particle. The sparticle spectra may become rather different from those
in other mechanisms for suppressing the FCNC processes and would give
distinctive signatures in future experiments.

\vspace*{5mm}
\subsection*{Acknowledgments}

The authors would like to thank to K.~Kurosawa, N.~Maekawa and
M.~Nojiri for helpful discussions and comments.


\end{document}